\documentclass[final]{siamltex1213}

\usepackage[latin1]{inputenc}
\usepackage{amsmath}
\usepackage{amsfonts}
\usepackage{amssymb}
\usepackage{graphicx}
\usepackage{relsize}
\usepackage{dsfont}
\usepackage{subfig}
\usepackage{float}  
\usepackage{todonotes}
\usepackage{multirow}
\usepackage{bm}

\title{A mechanism for cell motility by active polar gels\thanks{WM and AV acknowledge support from the German Science Foundation through grant Vo899/11. We further acknowledge support from the European Commission within FP7-PEOPLE-2009-IRSES PHASEFIELD and computing resources at JSC through grant HDR06.}} 

\author{W. Marth\thanks{Institut f\"ur Wissenschaftliches Rechnen, TU Dresden, 01062 Dresden, Germany 
({\tt wieland.marth@tu-dresden.de})} \and S. Praetorius\thanks{Institut f\"ur Wissenschaftliches Rechnen, TU Dresden, 01062 Dresden, Germany ({\tt simon.praetorius@tu-dresden.de})} \and A. Voigt\thanks{Institut f\"ur Wissenschaftliches Rechnen, TU Dresden, 01062 Dresden, Germany ({\tt axel.voigt@tu-dresden.de})}}

\begin{document}
\maketitle
\slugger{siap}{xxxx}{xx}{x}{x--x}

\begin{abstract}
We analyse a generic motility model, with the motility mechanism arising by contractile stress due to the interaction of myosin and actin. A hydrodynamic active polar gel theory is used to model the cytoplasm of a cell and is combined with a Helfrich-type model to account for membrane properties. The overall model allows to consider motility without the necessity for local adhesion. Besides a detailed numerical approach together with convergence studies for the highly nonlinear free boundary problem, we also compare the induced flow field of the motile cell with that of classical squirmer models and identify the motile cell as a puller or pusher, depending on the strength of the myosin-actin interactions.
\end{abstract}




\pagestyle{myheadings}
\thispagestyle{plain}
\markboth{{\sc W. Marth}, {\sc S. Praetorius} \and {\sc A. Voigt}}  {\sc A mechanism for cell motility by active polar gels}
 
\newcommand{\defeq}{\overset{\text{def}}{=}}
\newcommand{\inabla}{\mathlarger{\blacktriangledown}}%
\newcommand{\iDelta}{\mathlarger{\blacktriangle}}%
\newcommand{\dx}{\, \mathrm{d}\mathbf{x}}
\newcommand{\eps}{\varepsilon}
\newcommand{\ds}{\, \mathrm{d}s}

\section{Introduction}

Living cells move themselves around using different strategies, well adapted to their environment. A full understanding of the mechanisms behind cell motility is still missing but remains central for many biological and biomedical processes. Various generic mechanisms have been proposed to describe motility in different situations. Many eukaryotic cells for example move using a crawling motion. Here, motility results mainly from polymerization and depolymerization of actin filaments. The underlying treadmilling process, if combined with local adhesion of the cell on a substrate, leads to macroscopic motion. The treadmilling process and the associated crawling motion have been studied from a microscopic point of view, see e.g. \cite{Ziebertetal_JRSI_2012,Doubrovinskietal_PRL_2011} and \cite{Jilkineetal_PLOSCB_2011} for a review on existing mathematical models. Continuum models, which allow for spatial and temporal resolution, have been considered for such a crawling motility mechanism in \cite{Shaoetal_PRL_2010,Elliottetal_JRSI_2012,Shaoetal_PNAS_2012,Marthetal_JMB_2013,Ziebertetal_PONE_2013}. All these approaches use a reaction-diffusion system along the cell membrane and/or within the cytoplasm to effectively account for actin polymerization and combine it with a mechanical or hydrodynamic model for cell dynamics. This allows to describe the morphology and evolution of eukaryotic cells and link it to realistic signaling networks, as e.g. considered in \cite{Raetzetal_JMB_2012,Marthetal_JMB_2013}. 

Other motility mechanisms are less explored, but necessary in situations in which local adhesion is less evident, such as for cells moving in martigels \cite{Poinclouxetal_PNAS_2011,Hawkinsetal_BPJ_2011} or freely swimming microorganisms. We here consider a motility mechanism arising by contractile stress due to the interaction of myosin and actin. Microscopically, myosin motor complexes use the energy from ATP hydrolysis to grab on neighboring actin filaments and exert stress. This process is also known for eukaryotic cells, where it shapes the rear of the cell, but it can also lead to motility itself. Here, the exerted stress is contractile and leads to a microscopic quadrupole flow around the myosin-actin complexes. A hydrodynamic active polar gel theory is developed to model these phenomena on a continuum level, see \cite{Kruseetal_PRL_2000,Kruseetal_PRL_2004,Kruseetal_EPJE_2005}. If considered in a confinement, a splayed polarization of the filaments can occur and has already been used as a route to 
motility \cite{Tjhungetal_PNAS_2012,Giomietal_PRL_2014,Whitfieldetal_EPJE_2014}. All these studies consider a droplet. In the first case, with a surface tension using a numerical approach based on hybrid lattice Boltzmann simulations, in the second, the same setting is considered using a stream-function finite difference scheme and in the third a droplet of fixed shape is considered using an analytic description. 

We will here extend the approach to include also bending properties of a cell membrane, which, however, turns out to be of less relevance for the motility mode within the considered parameter regime. The focus of the paper is a detailed computational study of the motility mechanism due to myosin-actin interactions. We explain the used model, which is here formulated in a phase field description, demonstrate thermodynamic consistency of the overall model (without the active components), consider an adaptive finite element discretization in space and a semi-implicit time discretization for the system of equations and show convergence studies for critical parameters. As the considered motility mode results from a physical instability, a stable numerical discretization is essential for a detailed analysis. The simulation code is used to demonstrate the robustnes of the motility mechanisms and detailed parameter studies are provided to contribute to a better understanding of the mechanisms behind cell motility 
for environments without local adhesion. We also analyze the flow field induced by the motile cell and compare it with a squirmer model, which allows to identify the motile cell as a puller or pusher, 
depending on the strength of the myosin-actin interactions. We further discuss possible extensions of the model, e.g. combinations of myosin-actin interactions with actin polymerization. All simulations are restricted to 2D. The described model can also be used for 3D cell motility, where the myosin-actin interactions are assumed to dominate and treadmilling only plays a minor role. However, computational studies require an adequate preconditioner/solver for the system and its development is still current research. As already exemplarily shown in \cite{Tjhungetal_PNAS_2012} the motility mode remains persistent in 3D and we expect a similar robustness of the instability. However, a quantitative comparison of critical parameters, as well as comparisons with fluid flow measurements of moving cells will require computational intensive 3D simulations.

\section{Mathematical model}

The used model is an extension of the considered approach in \cite{Tjhungetal_PNAS_2012} and provides a generic route to study individual processes leading to cell motility. We will focus here on myosin-actin interactions as a source for cell motility. We review the equations and highlight the modifications.

\subsection{Energy}

We consider the free energy of the system
\begin{equation}
 E(\mathbf{P}, \phi, \mathbf{u})=E_\mathbf{P} + E_S + E_{kin} 
\end{equation}
which consists of the energy of the filament network $E_\mathbf{P}$ in the cytoplasm of the cell $\Omega_{cp}(t)$, described by an orientation field $\mathbf{P}$, which is the mesoscopic average orientation of the actin filaments, the surface energy $E_S$ of the cell membrane $\Gamma(t)$, described by a phase field variable $\phi$ and the kinetic energy $E_{kin}$ inside and outside of the cell, characterized by the velocity $\mathbf{u}$. For the sake of simplicity, we consider in the derivation equal density $\rho$ and viscosity $\eta$ for the cytoplasm and the fluid outside $\Omega_{out}(t)$, which is considered as an isotropic  Newtonian fluid, so that 
\begin{align}
E_{kin}=\frac{\rho}{2}\int_\Omega \mathbf{u}^2 \dx,
\end{align} 
with $\Omega = \Omega_{cp}(t) \cup \Gamma(t) \cup \Omega_{out}(t)$. 

The phase field variable is chosen, such that $\phi \approx 1$ in the cytoplasm and $\phi \approx -1$ in the fluid outside. The cell membrane is implicitly defined by the zero level set of $\phi$. In  \cite{Tjhungetal_PNAS_2012} the cell has been considered as a droplet for which the surface energy reads 
\begin{align}
 E_{S,CH} & =\frac{3\sigma}{2\sqrt{2}}\int_\Omega \frac{\eps}{2}|\nabla \phi|^2+\frac{1}{\eps} W(\phi)\dx
\end{align}
where $W(\phi)=\frac{1}{4}(\phi^2-1)^2$ denotes the double-well potential, $\eps$ determines the interface thickness and $\sigma$ is the surface tension. We here also take bending energy of the cell membrane into account and use the Helfrich \cite{Helfrichetal_ZNFC_1973}, or modified Willmore energy in a phase-field approximation \cite{Duetal_NONL_2005,Hausseretal_JBB_2013}
\begin{align}
 E_{S,W} & = \frac{3b_{N}}{4\sqrt{2}} \int_\Omega\frac{1}{2\eps}\left(\eps \Delta\phi-\frac{1}{\eps}(\phi^2-1)(\phi+\sqrt{2}H_0\eps)\right)^2\dx
\end{align}
where $b_N$ denotes the bending rigidity and $H_0$ the spontaneous curvature. We will set $H_0=0$ for simplicity. If $\eps$ tends to zero $E_{S,CH} \rightarrow  \sigma\int_\Gamma \ds$ \cite{Modica_ARMA_1987} and $E_{S,W}\rightarrow b_N\int_\Gamma (H-H_0)^2\ds$ \cite{Bellettinietal_CVPDE_2005} with $H$ the mean curvature. We will consider the combination of both surface energies
\begin{align}
 E_S=E_{S,CH}+E_{S,W}
\end{align}
for which $\Gamma$-convergence for $\eps \rightarrow 0$ was shown in \cite{Roegeretal_MathZ_2006}.

\begin{figure}
\centering
\includegraphics[width=0.4\textwidth]{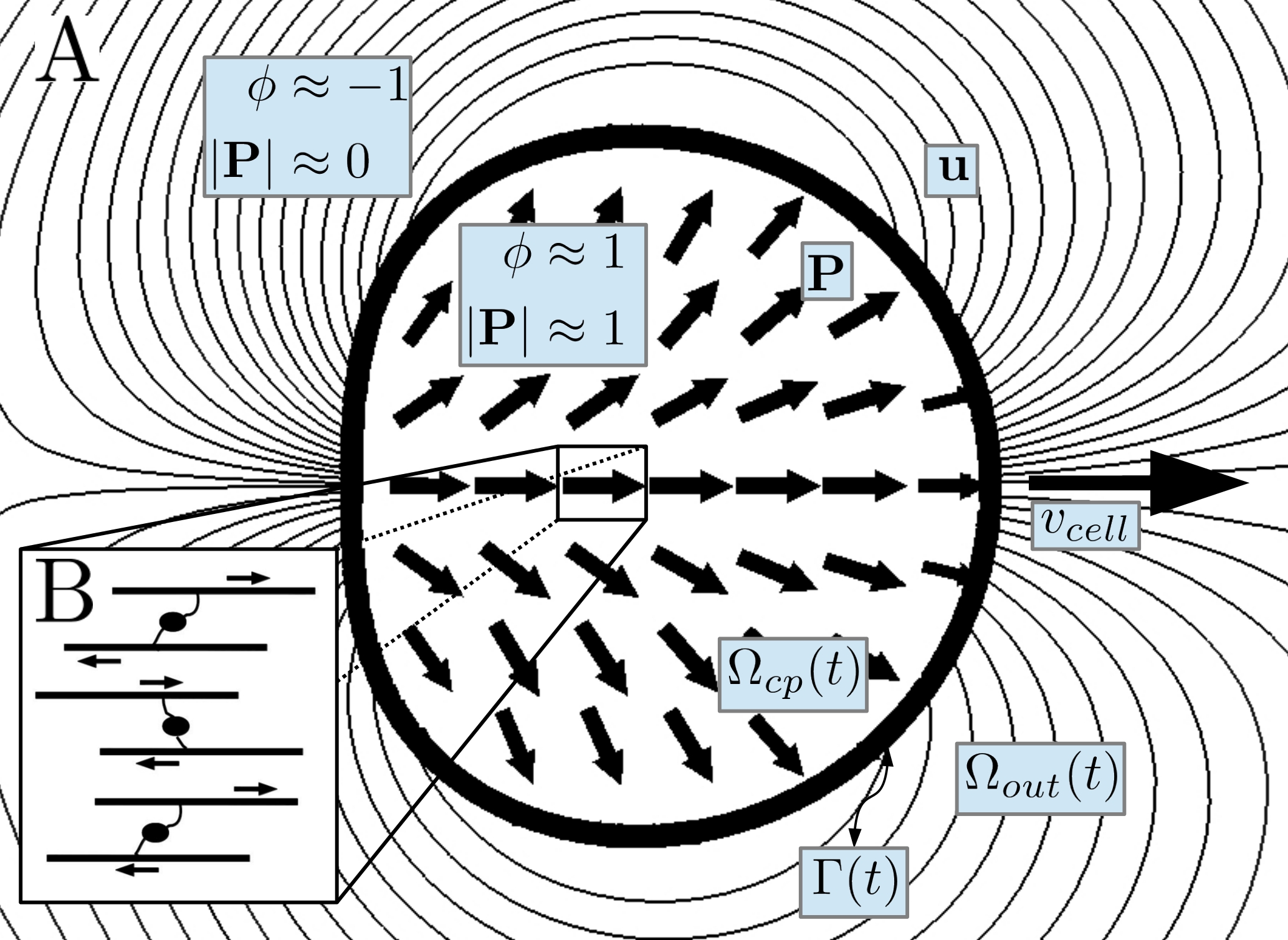}
\caption{(A) Schematic description for a moving cell. Shown is the splayed orientation field $\mathbf{P}$ in a motile steady state, with constant velocity $v_{cell}$ as well as the streamlines of the velocity profile $\mathbf{u}$ and the phase field $\phi$ with the cell membrane $\Gamma(t)$ corresponding to the zero-level set of $\phi$. (B) The orientation field serves as a model for the average aligned microscopic actin filaments which are connected by myosin motors.}
\label{fig:1}
\end{figure}

The energy of the filament network is given by \cite{Tjhungetal_PNAS_2012} 
\begin{align}
 E_P=\int_\Omega \frac{k}{2}(\nabla\mathbf{P})^2 + \frac{c_0}{4} |\mathbf{P}|^2 (- 2 \phi + |\mathbf{P}|^2) +\beta_0\mathbf{P}\cdot\nabla\phi\dx.\label{eq:energ}
\end{align}
The gradient term with the positive Frank constant $k$ is a simplification of a general distortion energy formulation from the theory of liquid crystals, with the assumption of the same value of the stiffness associated with splay and bend deformations, see e.g. \cite{deGennesetal_1993}.  Linking $\phi$ to the second term allows to restrict $\mathbf{P}$ to the cytoplasm: If $\phi<0$ the minimum is obtained for $|\mathbf{P}| = 0$ and thus the term does not contribute to the energy, and for $\phi>0$ the term forms a double-well with two minima with $|\mathbf{P}|=1$ and the form specified by the parameter $c_0$.  The last term in eq. (\ref{eq:energ}) guarantees for $\beta_0>0$ that $\mathbf{P}$ points outwards in normal direction to the cell boundary. This is expected to be of relevance for polymerization and depolymerization of actin filaments and used in \cite{Ziebertetal_JRSI_2012,Ziebertetal_PONE_2013,Loeberetal_SR_2015}, but for the here considered motility mode a strong preference of the orientation of $\
mathbf{P}$ at the cell boundary can not be seen. In \cite{Tjhungetal_PNAS_2012} it is argued that small $\beta_0$ values can resemble the effect of a weak external field. We will therefore consider both cases $\beta_0 = 0$ and $0 < \beta_0 \ll 1$ as in a more general approach with a combination of myosin-actin interactions and treadmilling $\beta_0>0$ will be required anyhow. Fig. \ref{fig:1} provides a schematic picture of the used variables.

Before we introduce the governing equations, we consider the energies in a non-dimensional form. We consider the characteristic values for space $\mathbf{x}=L \mathbf{\hat x}$, velocity $\mathbf{u} = U \mathbf{\hat u}$ and energy $E=\eta U L^2 \hat E$, with characteristic length $L$, characteristic velocity $U$ and fluid viscosity $\eta$. This yields a time scale $t=\frac{L}{U} \hat t$ and a pressure $p=\frac{\eta U}{L}  \hat  p$. We further define the constants $c_1 =\frac{c_0 L^2}{k}$ and $\beta=\frac{\beta_0 L}{k}$ and the dimensionless quantities:
\begin{itemize}
\item Reynolds number Re$\;= \frac{\rho U L}{\eta}$
\item Capillary number Ca$\;=\frac{ 2\sqrt{2}}{3}\frac{\eta U}{\sigma}$
\item Bending capillary number Be$\;=\frac{4\sqrt{2}}{3}\frac{\eta U L^2}{b_N}$
\item a polarity number Pa$\;=\frac{\eta U L }{k}$
\item an active force number Fa$\;= \frac{\eta U}{\zeta L}$,
\end{itemize}
where $\zeta>0$ describes a contractile and $\zeta<0$ an extensile stress. Dropping the ${\hat \cdot}$ notation we obtain the energies in a nondimensional form
\begin{align*}
E_{\mathbf{P}}&=\frac{1}{\text{Pa}}\int_\Omega \frac{1}{2}(\nabla\mathbf{P})^2+\frac{c_1}{4}|\mathbf{P}|^2 (-2\phi +|\mathbf{P}|^2)+\beta\mathbf{P}\cdot\nabla\phi\dx \label{eq:energNon} \\
 E_{S}  &=\frac{1}{\text{Ca}}\int_\Omega \frac{\eps}{2}|\nabla \phi|^2+\frac{1}{\eps} W(\phi)\dx + \frac{1}{\text{Be}} \int_\Omega\frac{1}{2\eps}\left(\eps \Delta\phi-\frac{1}{\eps}(\phi^2-1)\phi\right)^2 \!\!\! \dx \\
E_{kin}&=\frac{\text{Re}}{2}\int_\Omega \mathbf{u}^2 \dx
\end{align*}
which are used in the following.

\subsection{Governing equations}

The equations are based on \cite{Tjhungetal_PNAS_2012}. We denote the variational derivative or chemical potential of the orientation field and the phase field by $\mathbf{P}^\natural=\frac{\delta E}{\delta \mathbf{P}}$ and $\phi^\natural=\frac{\delta E}{\delta \phi}$, respectively.

\subsubsection{Orientation field equation}

The orientation field equation considers a polar liquid crystal theory combined with generalized hydrodynamics, see e.g.  \cite{Martinetal_PRA_1972,Brandetal_PRE_2006} and e.g. \cite{Ramaswamy_ARevCMP_2010,Marchettietal_RevModPhys_2013,Menzel_PR_2015} for a review, and is given by 
\begin{equation}
\partial_{t}\mathbf{P} + (\mathbf{  u} \cdot  \nabla)\mathbf{P} +  \mathbf{\Omega} \cdot \mathbf{P} =  \xi \mathbf{  D} \cdot \mathbf{P} - \frac{1}{\kappa}\mathbf{P}^\natural \label{eq:orient}
\end{equation}
where the left hand side is the co-moving and co-rotational derivative where the vorticity tensor defined as $\mathbf{\Omega}=\frac{1}{2}(\nabla \mathbf{u}^\top-\nabla \mathbf{u})$ takes rotational effects from the flow field into account, where $\nabla \mathbf{u}=(\partial_j u_i)_{(i,j)}$. The deformation tensor $\mathbf{D}=\frac{1}{2}(\nabla \mathbf{u}+\nabla \mathbf{u}^\top)$ and the nondimensional constant $\xi$ relates the coupling between the orientation field and the flow field and describes the alignment on $\mathbf{P}$ with the flow, where $\xi>0$ for rod-like and $\xi<0$ for oblate cells. Furthermore, $\kappa=\eta_{rot}/ \eta$ is a scaling factor between rotational and dynamic viscosity. The nondimensional chemical potential reads
\begin{align}
\mathbf{P}^\natural & = \frac{1}{\text{Pa}}\left(-c_1\phi \mathbf{P} + c_1 \mathbf{P}^2 \mathbf{P} -  \Delta \mathbf{P}+\beta \nabla  \phi\right). \label{eq:orientMob}
\end{align}

\subsubsection{Phase field equation}

We consider the phase field as an implicit representation of the cell surface and consider a regularized advection equation for the phase field variable $\phi$ with the advected velocity given by the fluid velocity $\mathbf{u}$. The introduced diffusion term is scaled with a small mobility coefficient $\gamma>0$. The evolution equation reads
\begin{equation}
 \partial_t \phi + \nabla \cdot \left(\mathbf{u} \phi \right) = \gamma \Delta \phi^\natural \label{eq:phase}
\end{equation}
with nondimensional chemical potential
\begin{align}
\phi^\natural = \frac{\delta E_\mathbf{P}}{\delta \phi} + \frac{\delta E_S}{\delta \phi} \label{eq:phaseMob}
\end{align}
with
\begin{align}
\frac{\delta E_\mathbf{P}}{\delta \phi}=\frac{1}{\text{Pa}}(-c_1 |\mathbf{P}|^2-\beta \nabla\cdot\mathbf{P}),
\end{align} 
which describes the influence of the orientation field and 
\begin{align}
\frac{\delta E_S}{\delta \phi}=\frac{1}{\text{Be}} \psi - \frac{1}{\text{Ca}} \mu,
\end{align}
which accounts for the bending and surface tension effects with
\begin{align}
\mu  & = \eps \Delta\phi-\frac{1}{{\eps}}(\phi^2-1)\phi,\\
\psi & = \Delta \mu-\frac{1}{{\eps}^2}(3\phi^2-1) \mu
\end{align}
introduced to write the higher order equation for $\phi$ as a system of 2nd order equations for $\phi, \mu, \psi$.

\subsubsection{Flow equations}

The physics of the flow are described by the Navier-Stokes equations
\begin{align}\begin{split}\label{eq:ns_1}
\text{Re}(\partial_t \mathbf{u} + (\mathbf{u}\cdot \nabla)\mathbf{u}) + \nabla p & =  \nabla\cdot \boldsymbol{\sigma}\\
\nabla\cdot\mathbf{u} &  = 0,
\end{split}
\end{align}
with hydrodynamic stress tensor $\boldsymbol{\sigma}=\boldsymbol{\sigma}_{viscous}+\boldsymbol{\sigma}_{active}+\boldsymbol{\sigma}_{dist}+\boldsymbol{\sigma}_{ericksen}$. The viscous stress is
\begin{align}
\boldsymbol{\sigma}_{viscous}= \mathbf{D}.
\end{align}
The active stress is
\begin{align}
\boldsymbol{\sigma}_{active}=\frac{1}{\text{Fa}} \tilde\phi \mathbf{P}\otimes \mathbf{P}
\end{align}
which describes the phenomenologically introduced activity \cite{Ramaswamy_ARevCMP_2010, Tjhungetal_SM_2011}, with $\tilde{\phi}=0.5(\phi+1)$ denoting the rescaled phase-field function, which serves as an approximation of a characteristic function for $\Omega_{cp}(t)$, with $\tilde{\phi}\approx 1$ in $\Omega_{cp}(t)$ and $\tilde{\phi}\approx 0 $ in $\Omega_{out}(t)$. The third term which describes the stress coming from the distortions of the filaments, reads
\begin{align}
\boldsymbol{\sigma}_{dist}=\frac{1}{2}(\mathbf{P}^\natural \otimes\mathbf{P} -\mathbf{P}\otimes \mathbf{P}^\natural) + \frac{\xi}{2}(\mathbf{P}^\natural\otimes \mathbf{P} + \mathbf{P} \otimes\mathbf{P}^\natural).
\label{eq:stress_dist}
\end{align}
For the Ericksen stress we consider the divergence to be defined through
\begin{align}
\nabla\cdot \boldsymbol{\sigma}_{ericksen}=\phi^\natural\nabla\phi + \nabla\mathbf{P}^T\cdot \mathbf{P}^\natural, \label{eq:sigma_ericksen}
\end{align}
which describes the stress coming from the cell surface as well as from the filaments as a result of their energy minimizing behavior, see \cite{Furthaueretal_NJP_2012,Catesetal_LNSI_2012}. This term also follows for the considered case $E_S = E_{S,CH}$ from the  explicit form used in \cite{Tjhungetal_PNAS_2012}.

\subsubsection{Initial and boundary conditions}

We consider a cell in a canal and take a rectangular domain $\Omega$. We assume periodic boundary conditions on the left and right boundary for all variables. At the upper and lower boundary we use homogeneous Neumann boundary conditions: $\nabla\mathbf{P}\cdot\mathbf{n}=\nabla\mathbf{P}^\natural\cdot\mathbf{n}=\mathbf{0}$, and  $\nabla\mu\cdot\mathbf{n}=\nabla\psi\cdot\mathbf{n}=0$ as well as Dirichlet boundary conditions $\mathbf{u}=0$ and $\phi=-1$. The initial condition for $\phi$ is the implicitly described initial cell shape $\phi = \tanh(r/(\sqrt{2} \eps))$, with $r$ the signed distance function to the membrane $\Gamma(0)$ and for $\mathbf{P}$ we apply an equal aligned filament network $\mathbf{P}=(P_1,P_2)^\top+\bm{\mathit{\delta}}$, where $\bm{\mathit{\delta}}$ is a vector-valued random number generated following an uniform distribution on the interval $[-0.05,0.05]$ in order to break the symmetry. For all simulations we start with a circular cell with the radius $R=5$ which is placed in the center 
of $\Omega=[0,160]\times[0,40]$. The initial condition for the orientation field is $\mathbf{P}=(1,0)^\top+\bm{\mathit{\delta}}$. 

\subsubsection{Material parameters}

We consider the following material parameters, see Tab. \ref{tab1}, which are adapted from \cite{Tjhungetal_PNAS_2012,Marthetal_JMB_2013} and the references therein. The low Reynolds number allows to restrict the flow equation to a Stokes system.

\begin{table}[ht]
\centering
\begin{tabular}{|l|l|l|}
\hline
Symbol & Description & Value \\
\hline \hline
$L$ & characteristic length & $10^{-6}$ m \\
$U$ & characteristic velocity& $10^{-6}$ m$/$s \\
$\rho$ & fluid density & $10^3$ kg$/$m$^{3}$ \\
$\eta$ & dynamic viscosity of the fluid & $2\cdot 10^{3}$ Pa$\,$s\\
$\sigma$ & surface tension           & $0.0188$ N$/$m\\
$b_N$ & bending rigidity & $1.26\cdot 10^{-14}$ N$\,$m\\
$k$ & Frank constant & $2\cdot 10^{-9}$ N, \cite{Tjhungetal_PNAS_2012, deGennesetal_1993} \\
$\xi$ & shape factor           & $1.1$, \cite{Tjhungetal_PNAS_2012} \\
$\eta_{rot}$ & rotational viscosity & $3.3\cdot 10^3$ Pa$\,$s, \cite{Tjhungetal_PNAS_2012} \\
$\zeta$ & activity parameter & $2 \cdot 10^3$ N$/$m$^2$, \cite{Whitfieldetal_EPJE_2014}\\
$\eps$ & boundary layer parameter & 0.21\\
$\gamma$ & mobility &  $0.025$\\
$c_1$ & double well parameter for $\mathbf{P}$ & 5 \\
$\beta$ & forcing normal direction of $\mathbf{P}$ at interface & 0, 0.005, 0.05\\
\hline
\end{tabular} 
\caption{
Material parameters of the system. For the given values we obtain the following characteristic numbers Ca=$0.1$, Be=$0.3$, Pa=$1$, Fa=$1$ and Re=$5\cdot 10^{-13}$.}
\label{tab1}
\end{table}

\subsubsection{Analytical results and numerical treatment}

Neglecting all active terms, the proposed system of equations fulfill thermodynamic consistency. This is shown in Appendix \ref{app:1}. If we further neglect the orientation field ($\mathbf{P} = 0$), the model reduces to a phase field approximation used for vesicle-fluid interactions, see e.g. \cite{Duetal_2007,Duetal_2009,Alandetal_JCP_2014}.
Further neglecting the bending forces by considering only $E_S = E_{S,CH}$ we obtain "Model H" in the classification of \cite{Hohenbergetal_RMP_1977}. If $\epsilon$ tends to zero this special case converges to a two-phase flow problem with a jump condition for the fluid stress tensor $- p \mathbf{I} + \mathbf{D}$ and a continuity condition for the fluid velocity $\mathbf{u}$ at the interface, see e.g. \cite{Abels_ARMA_2009}. Even if this analysis cannot easily be carried over to the full system, the last condition is expected to hold and thus guarantees that fluid cannot flow through the membrane.  

The system of partial differential equations is discretized using the parallel adaptive finite element toolbox AMDiS \cite{Veyetal_CVS_2007,Witkowskietal_ACM_2015}. We further explore an operator splitting approach, allowing to solve the subproblems of the flow field, the orientation field and the phase-field evolution separately in an iterative process. In time, a semi-implicit discretization is used, which, together with an appropriate linearization of the involved non-linear terms, leads to a set of linear systems in each time step. Details are described in 
Appendix \ref{app:2}.

\section{Simulations}

\subsection{Motility due to contractile and extensile stress}
\begin{figure}[h]
\includegraphics[scale=0.8]{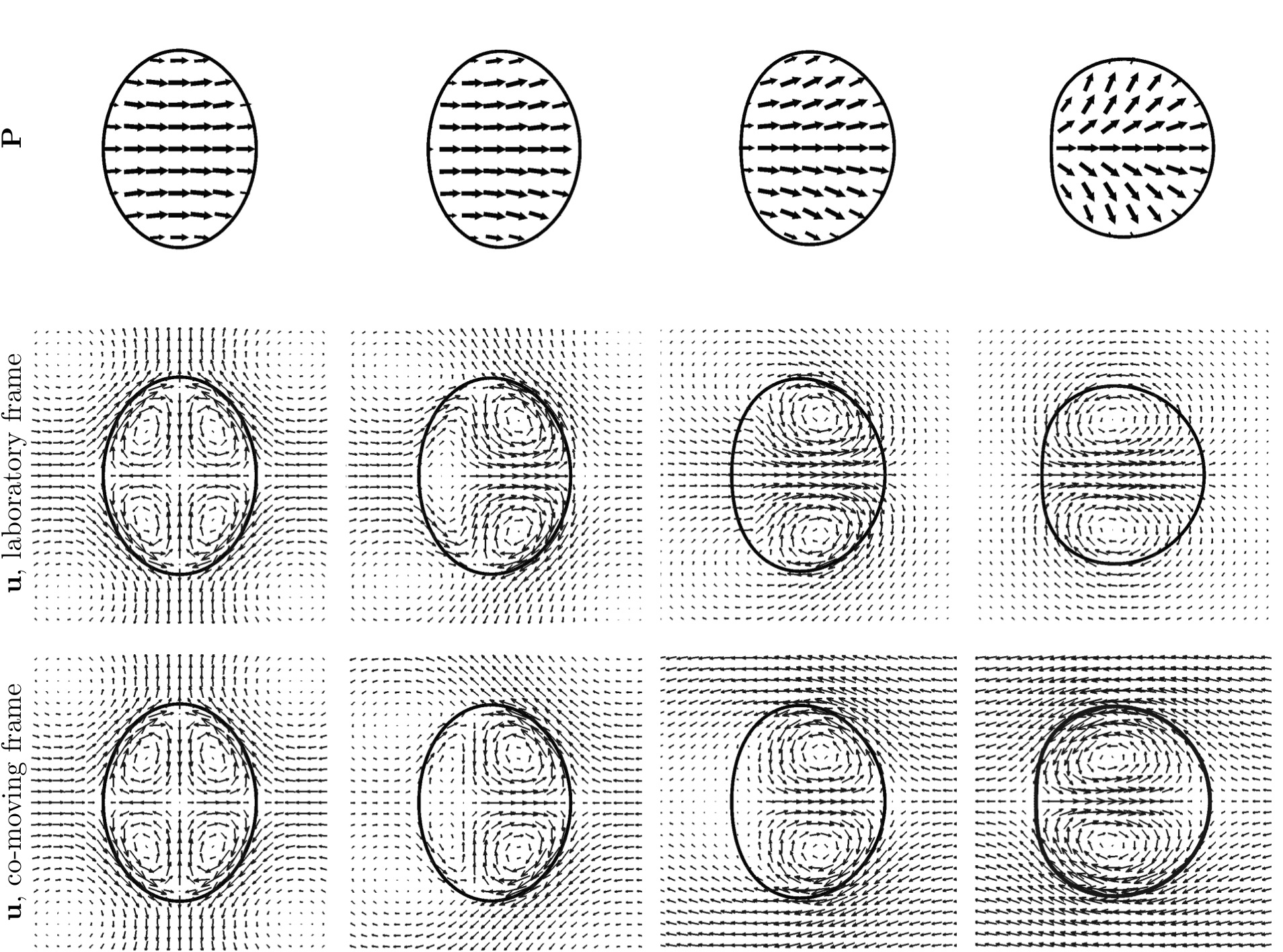}
\caption{Cell movement for contractile stress, movement to the right: first row - shape of the cell at different times evolving from left to right. Shown is the magnitude and the direction of the orientation field. second row - velocity field in a laboratory frame with different maxima: $|\mathbf{u}|=0.1$, $|\mathbf{u}|=0.12$, $|\mathbf{u}|=0.19$ and $|\mathbf{u}|=0.42$, which correspond to the cell speed $v_{cell}$ of $0$, $0.016$, $0.054$ and $0.125$ from left to right. third row - velocity field of the co-moving frame, i.e. $(u_1-v_{cell},u_2)^T$. The times $t$ shown are 100, 220, 250, 340, which correspond to seconds. The values used are from Table \ref{tab1} and we changed 1/Fa=$1.125$ and take $\beta=0$ (no explicit forcing for $\mathbf{P}$ to point outwards at the cell boundary).}
\label{fig9}
\end{figure}

As in \cite{Tjhungetal_PNAS_2012} motility can be achieved by means of a spontaneous splay deformation. It is a two-stage process, with an elongation of the cell as a consequence of a quadrupolar straining flow resulting from the active stress tensor $\boldsymbol\sigma_{active}$. The elongation stops, if the surface forces characterized by Ca and Be balance the active stress. The orientation field $\mathbf{P}$, which remains rather uniform during the elongation, starts to fluctuate, which induces a shear flow parallel to the orientation field and a spontaneously splay instability. The splayed configuration breaks the axial symmetry of the system and transforms the quadrupolar flow in a dipolar flow with two large vortices running across the cell, which has an influence on the cell shape and causes the cell to move with constant shape and at constant velocity along the symmetry axis, see Fig. \ref{fig9}.

\begin{figure}[h]
\includegraphics[scale=0.8]{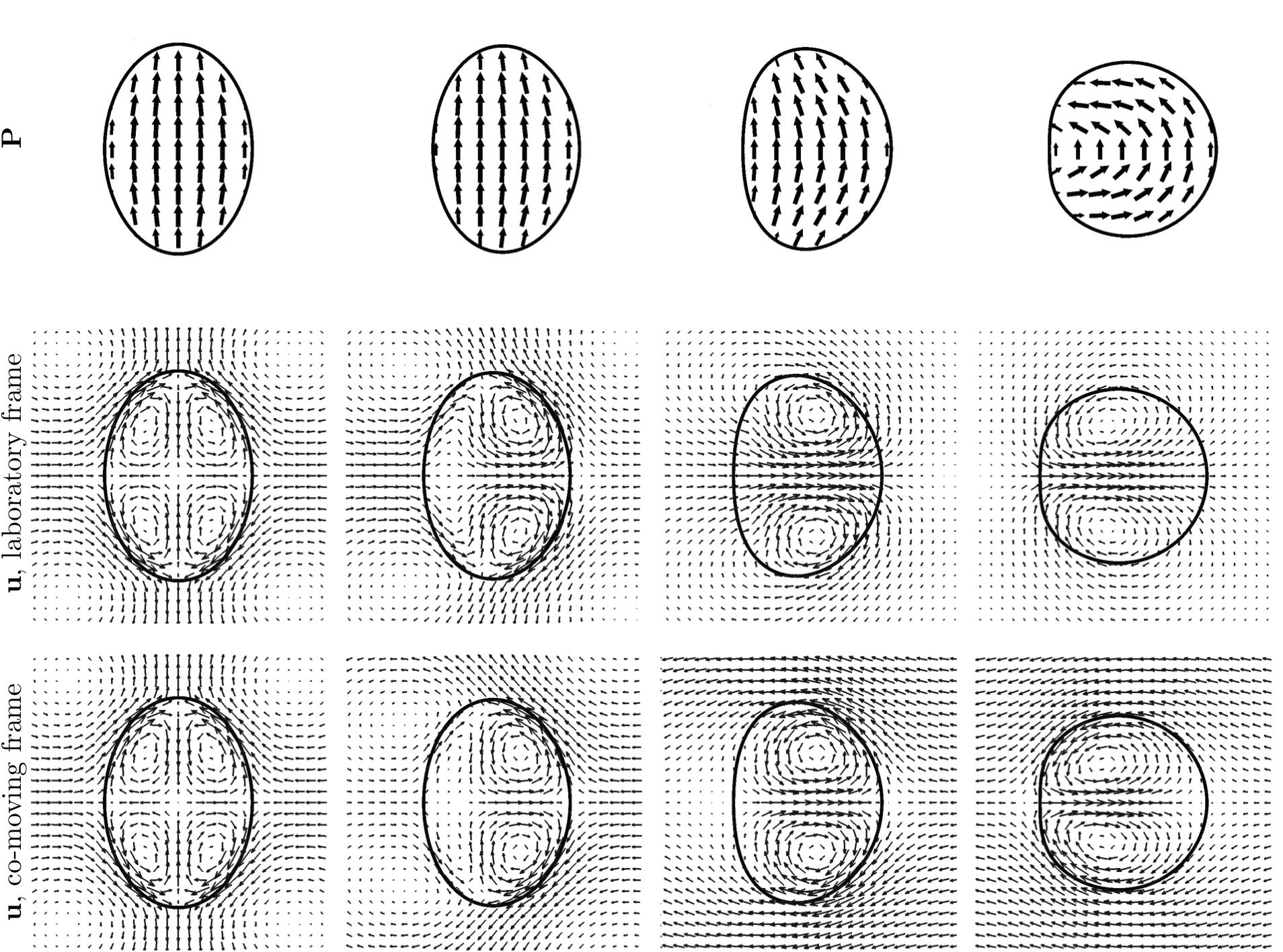}
\caption{Cell movement for extensile stress, movement downwards. first row - shape of the cell at different times evolving from left to right. Shown is the magnitude and the direction of the orientation field. second row - velocity field in a laboratory frame with different maxima: $|\mathbf{u}|=0.117$, $|\mathbf{u}|=0.138$, $|\mathbf{u}|=0.266$ and $|\mathbf{u}|=0.73$, which correspond to the cell speed $v_{cell}$ of $0$, $0.02$, $0.07$ and $0.16$ from left to right. third row - velocity field in a co-moving frame, i.e. $(u_1-v_{cell},u_2)^T$. Note that bend instabilities generate a moving direction normal to the initial direction of the orientation field. The times $t$ shown are 10, 80, 100, 170, again corresponding to seconds. The parameters are the same as in Fig. \ref{fig9}.}
\label{fig10}
\end{figure}

For completeness, we also demonstrate an example for cell motility due to extensile stress. Here, the vortices are reverse and the cell is stretched in the $x_1$-direction. Together with the active stress, which now generates a flow normal to the filaments, a bend instability occurs, describing an alignment of the filaments along the curved shape of the cell. This results in a downward motion, see Fig. \ref{fig10}. The only modification needed to achieve this, is 1/Fa=$-3/2$.

\begin{figure}[h]
\centering
\includegraphics[scale=0.8]{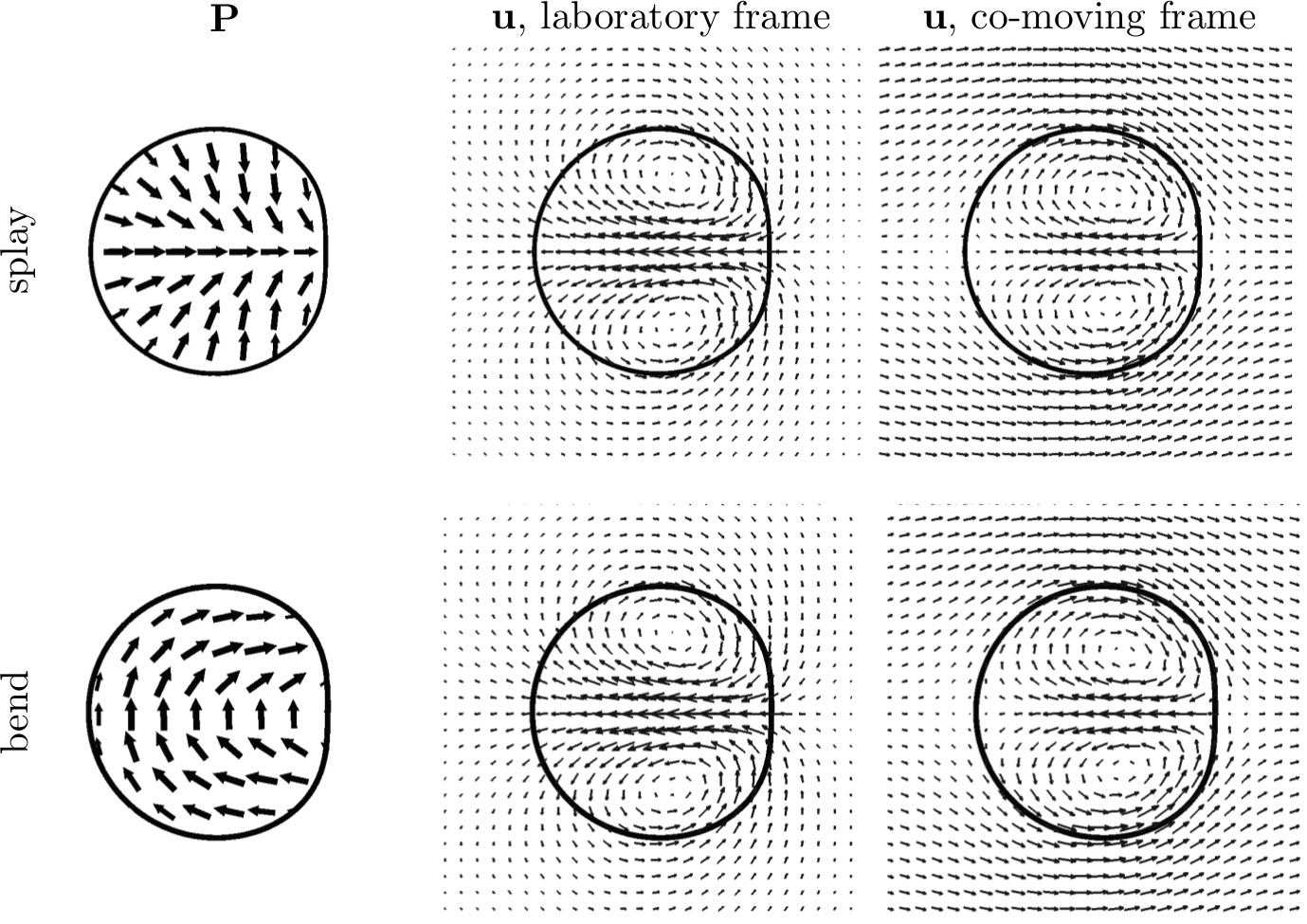}
\caption{Opposite instabilities:  Depending on the initial conditions as well as on the external effects the splay instability (first row) and bend instability (second row) draw a different pattern (left) and the cell moves in the opposite direction, to the left and upwards, respectively.}
\label{fig11:opposite}
\end{figure}

The shape and the direction of both instabilities depend on the initial conditions as well as small disturbances due to external influences. Fig \ref{fig11:opposite} shows the opposite splay instability (first row) and the opposite bend instability (second row). Although the cell moves in the contrary direction the velocity profile has a similar shape as before.

All these results qualitatively agree with \cite{Tjhungetal_PNAS_2012}. We now turn to more quantitative comparisons and test the robustness of the instabilities.

\subsection{Onset of motility}

\begin{figure}[!ht]
\center
\includegraphics[scale=0.7]{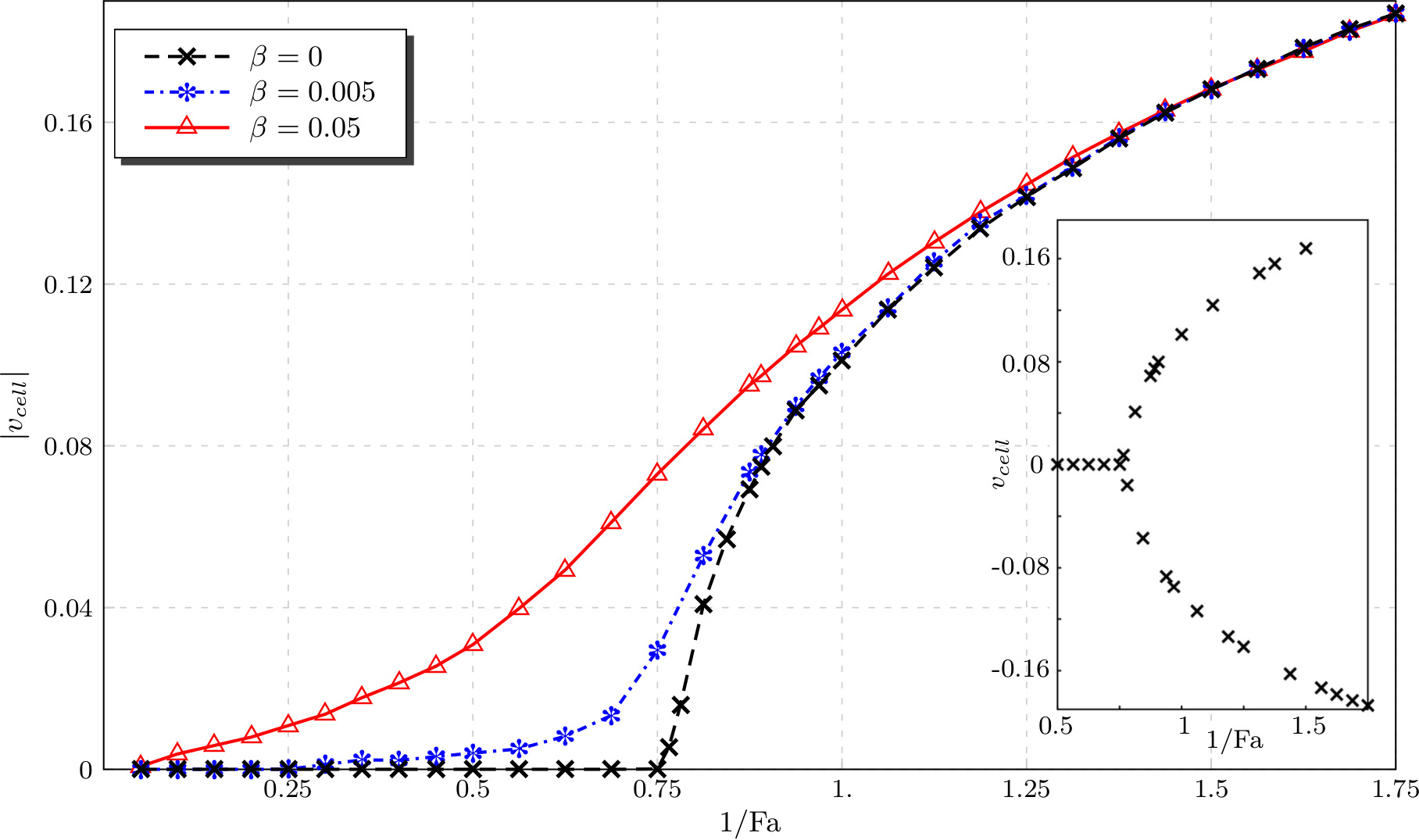}
\caption{Bifurcation diagram showing the symmetry breaking from a stationary state to a splayed and moving state for increasing 1/Fa. For $1/\text{Fa} < 1/\text{Fa}_{crit}$ the cell remains stationary and for $1/\text{Fa} > 1/\text{Fa}_{crit}$ the cell is moving, shown is the absolute value of $v_{cell}$. This transition is smoothed out for $\beta > 0$. The inlet shows both branches of the diagram with opposite velocities which occur only for the case $\beta = 0$.}
\label{fig12}
\end{figure}

In any case, motility is only possible if the strength of the myosin-actin interactions exceeds a critical value. We obtain a critical activity parameter $1/\text{Fa}_{crit} \approx 0.75$. Below 1/Fa$_{crit}$ no instability occurs and the cell does not move. This is at least the case for $\beta = 0$ and in qualitative agreement with \cite{Tjhungetal_PNAS_2012}. The bending capillary number Be does not influence the behaviour within the considered parameter regime. However, a quantitative comparison with the results in \cite{Tjhungetal_PNAS_2012}, where $1/\text{Fa}_{crit} \approx 0.5$ is measured, cannot be achieved as not all parameters used in \cite{Tjhungetal_PNAS_2012} are known and the critical value turns out to be highly sensitive to various parameters, which will be analyzed below. Fig. \ref{fig12} shows the upper branch of the bifurcation diagram separating a stationary state from a splayed and moving state by plotting the constant velocity of the cell. For $\beta>0$ the transition to a immotile 
cell is smoothed out. We no longer have a sharp transition and observe motility also below 1/Fa$_{crit}$, again in agreement with \cite{Tjhungetal_PNAS_2012}.

The onset of the instability and the time required to reach a constant shape moving with constant velocity depends on the used parameters. As stronger the myosin-actin interactions, as faster this shape is reached. This effect is most pronounced for $\beta = 0$ and decreases for $\beta > 0$. The time to reach a constant shape moving with constant velocity also depends on membrane properties of the cell. While the bending capillary number Be only plays a minor role in the considered parameter regime, the influence of the capillary number Ca is significant. The smaller the surface tension, the longer it takes to reach the desired shape. Again, this effect is less pronounced for $\beta > 0$.

\subsection{Convergence tests}

All obtained results are very sensitive to various parameters. The motility results from a splay or bend instability, which e.g. is heavily influenced by the elasticity of the filament network, related to the Frank constant $k$, which is here carefully chosen together with other physical parameters to observe the instability.  Due to this sensitivity on the physical parameters, we would like to consider the influence of numerical parameters on the described phenomena.

\begin{figure}[ht]
\begin{center}
\includegraphics[scale=0.8]{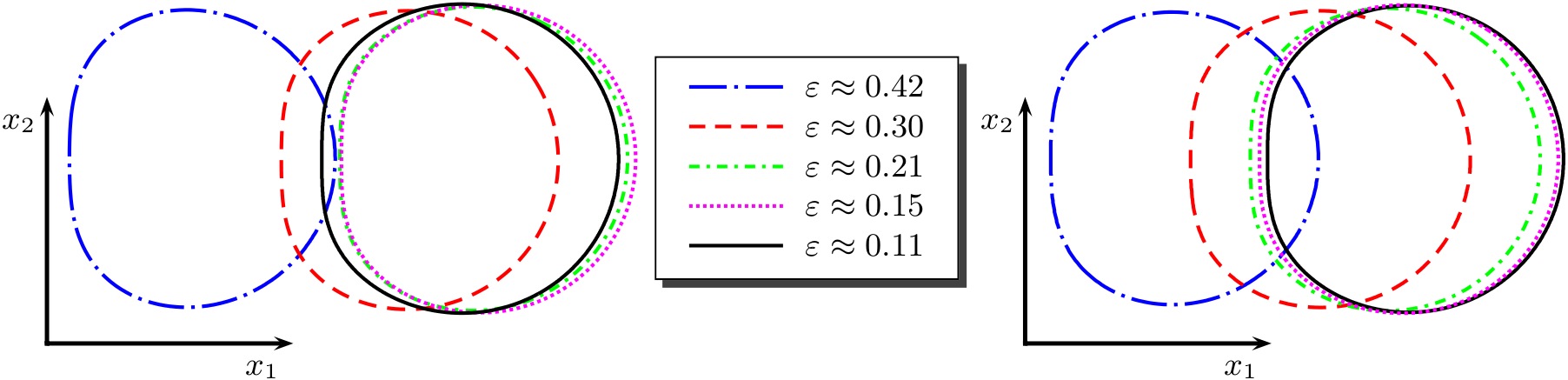}
\end{center}
\caption{Relative cell positions and cell shapes for for different interface thicknesses $\eps$ in case of $\beta=0$ (left) and $\beta=0.05$ (right) at time $t=300$. If the cell moves to the left (in case of $\beta=0$), we reflect the cell shape with respect to the $x_2$ axis of the initial center of mass.}
\label{fig11}
\end{figure}

We consider convergence tests. As we are primarily interested in cell motility, we first consider a parameter regime for which our cell becomes motile and moves with a constant shape and constant velocity. We consider the case of contractile stress and thus, movement in horizontal direction. We use shape and velocity for validation and measure the following quantities: 
\begin{itemize}
\item the $x_1$-coordinate of the center of mass, 
\[
x_{cm}= \frac{1}{|\Omega_{cp}|}\int_{\Omega_{cp}} x_1 \, \dx,
\]
$\mathbf{x} = (x_1,x_2)^\top$ and $|\Omega_{cp}|=\int_{\Omega_{cp}} 1 \, \dx$,
\item the mean velocity of the cell 
\[
u_{cell} = \frac{1}{|\Omega_{cp}|} \int_{\Omega_{cp}} u_1 \, \dx,
\] 
as an average of the $x_1$-component of the velocity in $\Omega_{cp}$, where $\mathbf{u}=(u_1,u_2)^\top$, and 
\item the circularity of the cell, which is defined as the quotient of the perimeter of an area-equivalent circle and the perimeter of the cell 
\[
c_{cell} = \frac{2}{B(\phi)}  \left(\int_{\Omega_{cp}} \pi \dx\right)^{1/2},
\] 
where $B(\phi)$ is the perimeter of the cell.
\end{itemize}
We used absolute values for all quantities and the following error norm: $\|e\|_2 = ( (\sum_{I} |q_{t,ref} - q_t|^2) / (\sum_I |q_{t,ref}|^2) )^{1/2}$, where $q_t$ is the temporal evolution of quantity $q$. The solution on the finest grid serves as reference solution $q_{t,ref}$. Tab. \ref{tab:convergence1} shows the relative error norms as well as the relative order of convergence (ROC) for the desired quantities if $\eps$ is reduced. We consider two cases $\beta=0$ and $\beta=0.05$. Together with $\eps$ we also refine the mesh size to guarantee the same number of grid points within the diffuse interface layer for all simulations and the time step to ensure the same relation between mesh size and time step. The time interval is $I =[0,500]$. Other parameters are obtained from Tab. \ref{tab1}. We see essentially first order convergence, the higher numbers in ROC are probably due to fortunate circumstances. Fig. \ref{fig11} show the shape and position for various $\eps$, visualizing the convergence and 
confirming the choice of $\eps = 0.21$ for the previous and further studies.

\begin{table}[h]
\centering
\begin{tabular}{|l|l|ll|ll|ll|}
\hline
        &       & \multicolumn{2}{c|}{center of mass $x_{cm}$} &  \multicolumn{2}{c|}{cell velocity $v_{cell}$} &  \multicolumn{2}{c|}{circularity $c_{cell}$}\\
        &$\eps$ & $\|e\|_2$        & ROC                       & $\|e\|_2$ & ROC                                & $\|e\|_2$ & ROC        \\    
\hline\hline
\multirow{4}{*}{$\beta=0$} &$0.42$ &  0.0600         &                          &   0.3988        &                              & 0.0398    &            \\
    &$0.30$ &  0.0177         &  3.5298                  &   0.1659        &   2.5316                     & 0.0314    &  0.6823          \\
 &$0.21$ &  0.0047         &  3.8355                  &   0.0787        &   2.1516                     & 0.0179    &  1.6157          \\
 &$0.15$ &  0.0028         &  1.4912                  &   0.0273        &   3.0575                     & 0.0061    &  3.1225           \\
\hline\hline
\multirow{4}{*}{$\beta=0.05$} &$0.42$ &  0.0569         &                          &   0.2575        &                              & 0.0430    &            \\
 &$0.30$ &  0.0298         &  1.8715                  &   0.1302        &   1.9691                     & 0.0316    &  0.8921          \\
 &$0.21$ &  0.0129         &  2.4195                  &   0.0511        &   2.6938                     & 0.0174    &  1.7122          \\
 &$0.15$ &  0.0025         &  4.7328                  &   0.0095        &   4.8514                     & 0.0059    &  3.1431           \\
\hline
\end{tabular} 
\caption{Relative error norms and convergence orders for critical parameters, upper part $\beta=0$ and lower part $\beta = 0.05$.}
\label{tab:convergence1}
\end{table}

\begin{figure}[ht]
\centering
\includegraphics[scale=0.8]{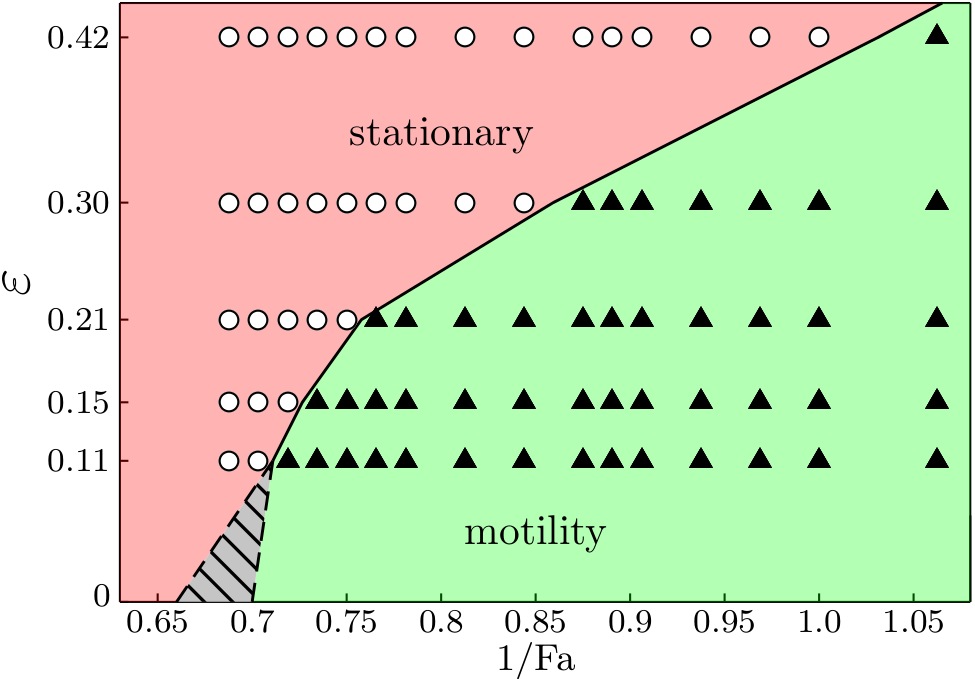}
\caption{Phase diagram distinguishing between stationary and motile state as function of 1/Fa and $\eps$. 1/Fa$_{crit}$ can be considered as a function of $\eps$ with the limiting value for $\eps\to 0$ presumably within the shaded region.}
\label{fig:zetacrit}
\end{figure}

\begin{figure}[ht]
\centering
\includegraphics[scale=0.8]{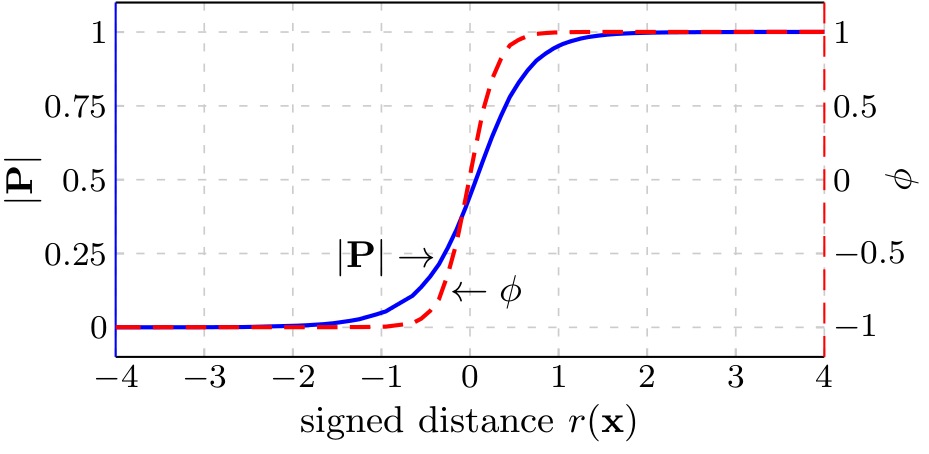}
\caption{1D cut of the phase field function $\phi$ and norm of the polarization field $\mathbf{P}$ for $\eps=0.3$}
\label{figDraw4c}
\end{figure}

The second test considers the onset of motility. How sensitive is the obtained critical parameter 1/Fa$_{crit}$ on $\eps$? The relation is shown in Fig. \ref{fig:zetacrit}. A deeper analysis of the interface profile, as shown for a 1D cut of a cell in Fig. \ref{figDraw4c} explains this dependency as $|\mathbf{P}|$ is slightly more smeared out than $\phi$. This has an influence on the active stress $\boldsymbol{\sigma}_{active}$. Its divergence is reduced at the interface for increasing $\epsilon$ and therefore a larger activity is needed to initiate the instability.

\subsection{Influence of different viscosities}

\begin{figure}[ht]
\centering
\includegraphics[scale=0.8]{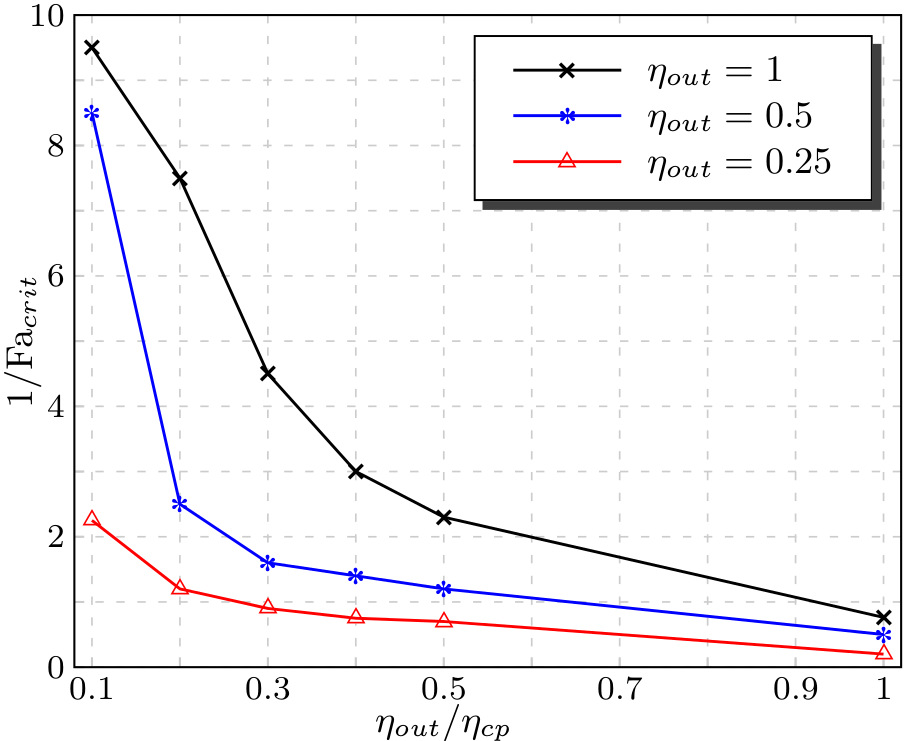}
\caption{Dependency of $1/Fa_{crit}$ on viscosity ratio between outside fluid and the cytoplasm. $\eta_{out} = \eta_{cp} = 1$ corresponds to the previously considered case.}
\label{figeta}
\end{figure}

Up to now we have considered equal density and viscosity for the cytoplasm and the fluid outside. The model can easily be extended to relax this restriction. We hereby follow a typical extension of "Model H", taking $\rho = \rho(\phi)$ and $\eta = \eta(\phi)$. As shown in \cite{Alandetal_JNMF_2012} the results for this approach are comparable to other more advanced approaches. In the following we only consider variations in $\eta$ and define $\boldsymbol{\sigma}_{viscous} = \tilde{\eta}(\phi) \mathbf{D}$ with an appropriate function $\tilde{\eta}(\phi)$ interpolating between $\eta_{out}$ and $\eta_{cp}$, which are rescaled dimensionless numbers corresponding to the viscosity in the fluid outside and the cytoplasm, respectively. Fig. \ref{figeta} shows the dependency of $1/Fa_{crit}$ on the values of $\eta_{out}$ and $\eta_{cp}$. Decreasing the viscosity, but keeping both values equal, leads to a reduction of the required activity for motility, but increasing the viscosity in the cytoplasm and keeping the 
viscosity in the outside fluid constant, in all cases, leads to an increase of the required activity. This can be explained by the necessity to induce a characteristic flow pattern in $\Omega_{cp}$ to induce the instability, which becomes harder to achieve for larger viscosities. 

The viscosity also has an influence on the cell velocity. The reached stationary velocity $v_{cell}$ increases if $\eta_{out}$ is reduced. For more realistic parameters, with an even larger ratio of $\eta_{out}/\eta_{cp}$ we thus expect faster moving cells. The slope of the corresponding bifurcuation branch, as in Fig. \ref{fig12}, above $1/Fa_{crit}$ is reduced if $\eta_{cp}$ is increased. The sharp transition to motility for $\beta = 0$ and the smoothed out transition for $\beta > 0$ remain. 

\section{Discussion}

\begin{figure}[t]
\centering
\includegraphics[scale=0.8]{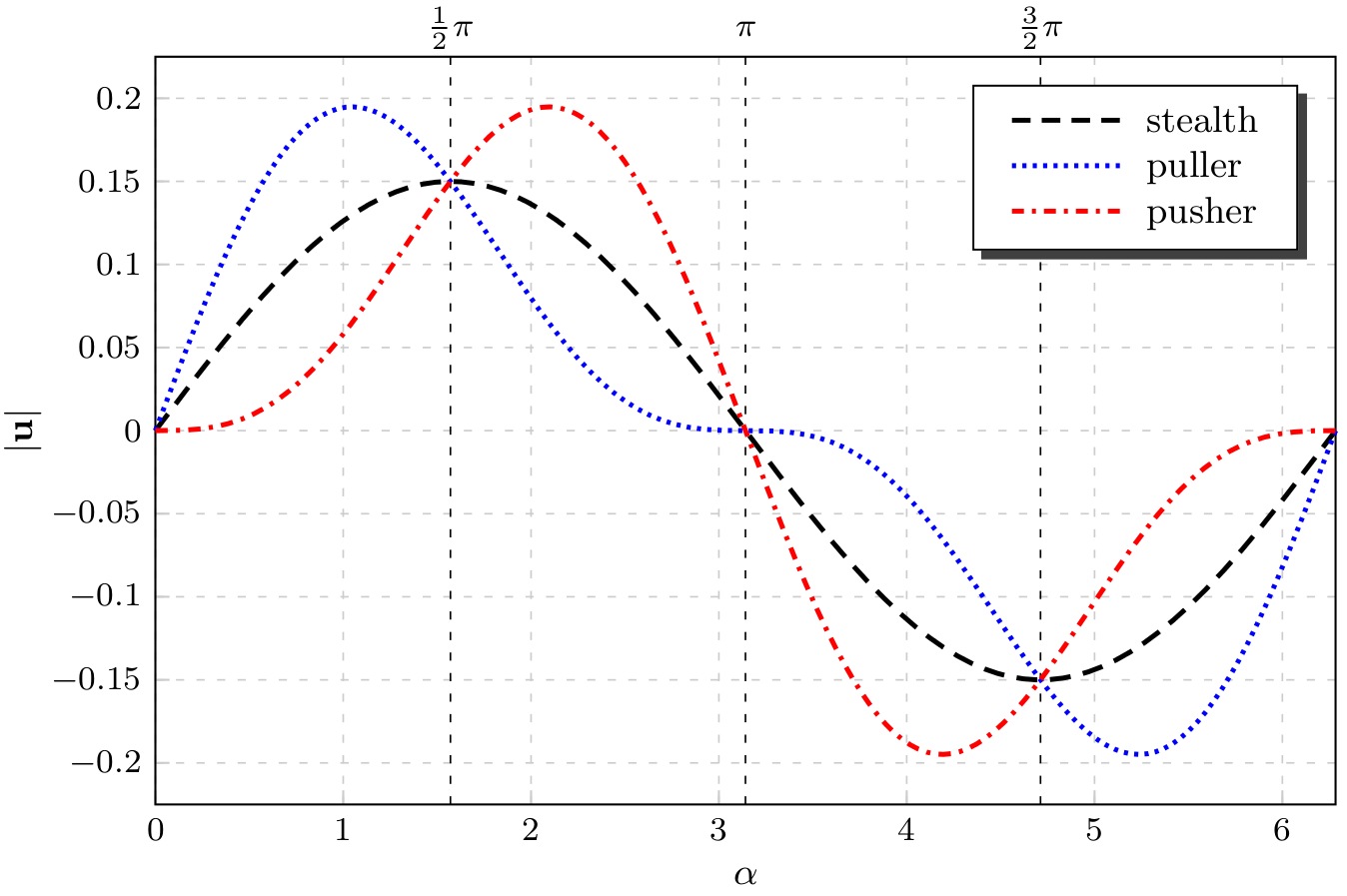}
\caption{Analytical solutions of the velocity profiles $\mathbf{u}_{T,squirmer}$ along the interface (Eq. \ref{eq:analyt_squirmer}), for different swimmer types.}
\label{fig:swimmer_gen}
\end{figure}

We already emphasized, that this model describes cell motility without adhesion. Can we relate the motility mode to any freely-swimming microorganism? In order to answer this question, we first compare the induced flow field with theoretical predictions for a squirmer model \cite{Lighthill_ARevFM_1969,Blakeetal_JFM_1971} and e.g. \cite{Molinaetal_SoftMatter_2013}. The surface tangential velocity for a circular squirmer in a co-moving frame in polar coordinates is given by
\begin{equation}
 \mathbf{u}_{T,squirmer}=n_1(\sin{\alpha}+m \sin{2\alpha}) \label{eq:analyt_squirmer}
 \end{equation}
where $n_1$ determines the velocity of the cell, whereas $m=\frac{n_2}{n_1}$ defines whether the swimmer is a pusher ($m<0$), a puller ($m>0$) or a neutral (stealth) swimmer ($m=0$), and $\alpha$ is the angle between the swimmers fixed swimming axis and the vector pointing to the surface. Figure \ref{fig:swimmer_gen} shows the surface tangential velocity for different swimmers, where we choose $n_1=0.15$ as well as $m=0$ (stealth), $m=0.5$ (puller) and $m=-0.5$ (pusher). The profiles significantly differ with the extrema in that part of the swimmer, which is responsible for the motion. In case of a puller it is the cell front $(0< \alpha< \pi/2)$ and $(3\pi/2 < \alpha< 2\pi)$, whereas as the pusher is driven by the rear, so the extrema appear for $ (\pi/2< \alpha< 3\pi/2)$. For a neutral swimmer the extrema are at $\pi/2$ and $3/2 \pi$. 

\begin{figure}[ht]
\centering
\includegraphics[scale=0.8]{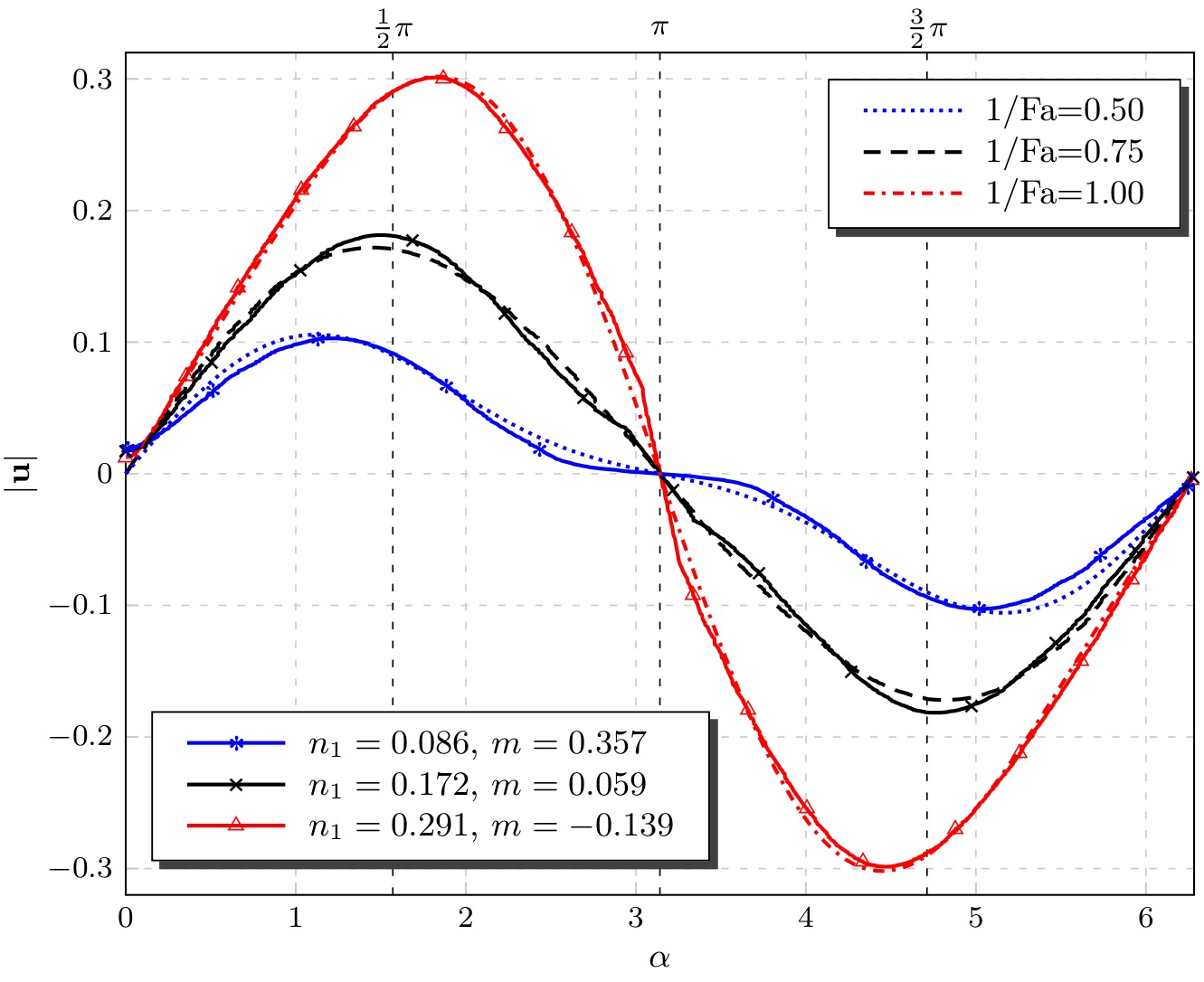}
\caption{velocity profiles $\mathbf{u}_{T}$ (dashed lines) and corresponding datafit (solid lines) for various parameters 1/Fa.}
\label{fig:swimmer_fit}
\end{figure}

\begin{figure}[H]
\centering
\includegraphics[scale=0.8]{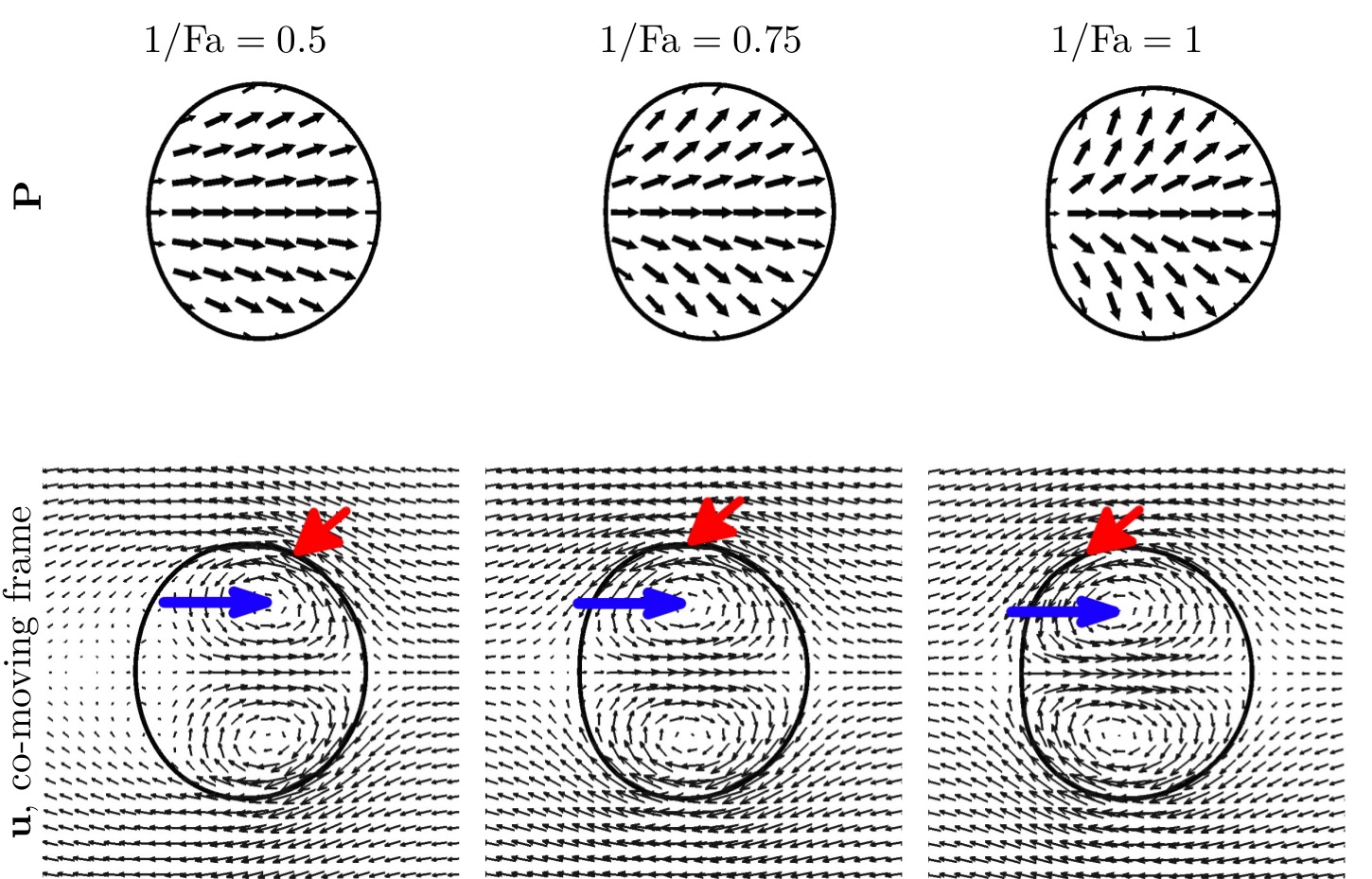}
\caption{Stationary shapes moving with constant velocity to the right for different 1/Fa: 0.5, 0.75, 1, from left to right. Polarization field (first row), velocity field in co-moving domain (second row). With increasing activity, the splay instability is enhanced, which moves the maximum of the velocity field along the interface (indicated by red angular arrow) from the front to the rear, visible also through the position of the vortices in the cell (indicated by blue arrow), which are located more towards the front for puller dynamics and more towards the rear for pusher dynamics.}
\label{fig:splay_swimmer}
\end{figure}

We now compare these results with our simulations. We therefore extract the surface tangential velocity in the co-moving frame from our simulations. We use a contractile stress and consider $\mathbf{u}_T=\left.(u_1-v_{cell},u_2)^\top\right|_{\phi(x,\tilde t)=0}$ for $\tilde t>0$ such that the stationary profile and velocity is reached. Figure \ref{fig:swimmer_fit} shows the profile for various parameters 1/Fa and $\beta = 0.05$. In comparison with the analytical results, we find puller dynamics for 1/Fa $\le 0.5$, similarities to neutral swimmers for 1/Fa = 0.75 and pusher dynamics for 1/Fa $\ge 1$. For $\beta = 0$ we qualitatively obtain the same results for 1/Fa $\geq$ 1/Fa$_{crit}$ and thus only pusher dynamics. The corresponding velocity profiles from the squirmer model are obtained from a data fit (see Figure \ref{fig:swimmer_fit}): $n_1=0.086$, $m=0.357$ (puller), $n_1=0.172$, $m=0.059$ (neutral) and $n_1=0.291$, $m=-0.139$ (pusher), respectively. Although we are comparing results for nearly circular 
shapes, see Fig. \ref{fig:splay_swimmer} for the corresponding stationary profiles, with that from analytic results for circular shapes, we observe a 
reasonable agreement.

\begin{figure}[ht]
\includegraphics[width=\textwidth]{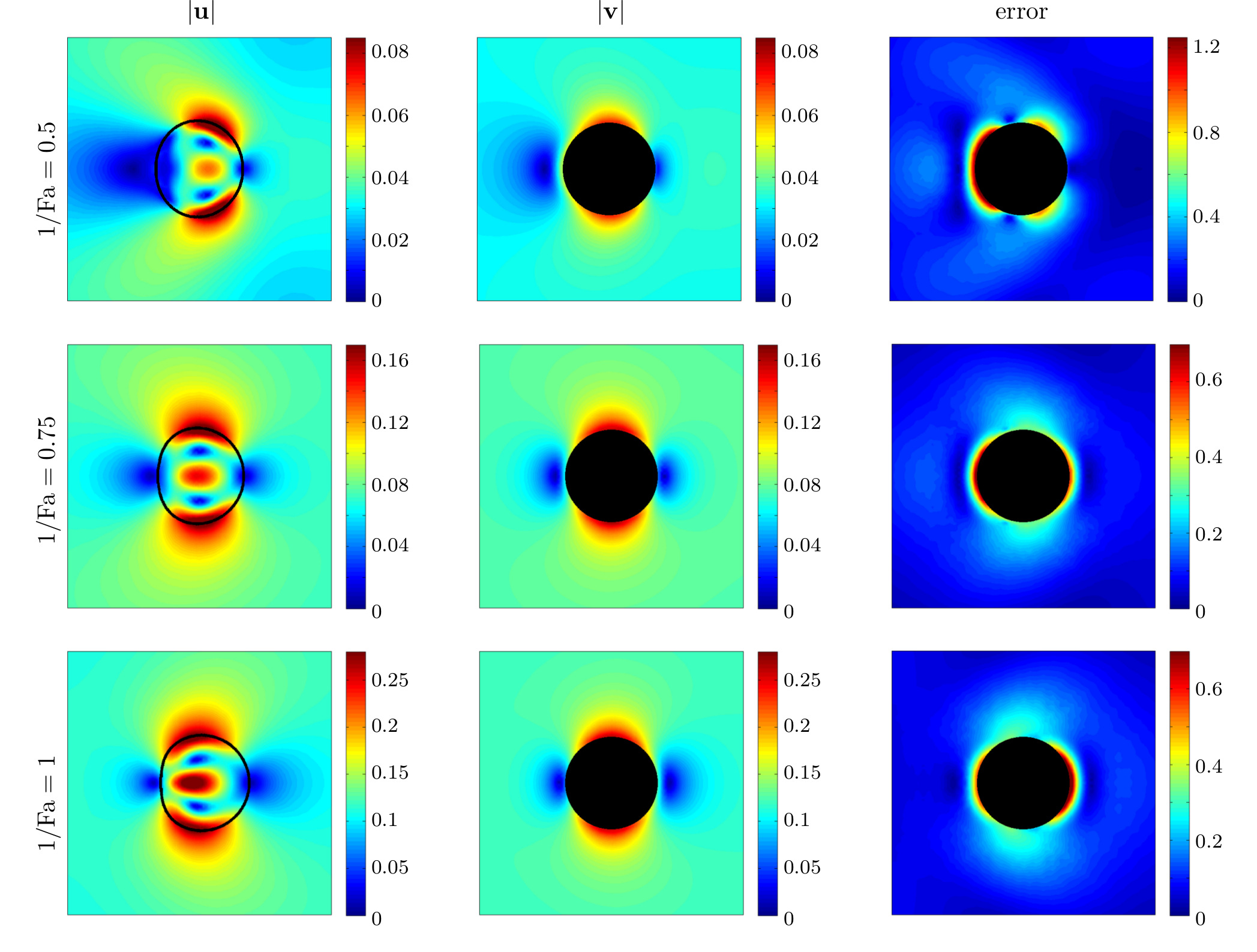}
\caption{Magnitude of the velocity profile of the numerical solution $|\mathbf{u}|$ (left), the fitted analytical solution $|\mathbf{v}|$ (middle) and the relative error (right) for 1/Fa=0.5 (first row), 1/Fa=0.75 (second row) and 1/Fa=1 (third row). For the analytical solution as well as the error analysis we approximated the cell shape by a circle, with radius $R=5$ obtained from the initial condition. The data fit indicates 1/Fa=0.5 as puller and 1/Fa=1 as pusher. For 1/Fa=0.75 we get a puller like velocity profile, where we expected a neutral swimmer, which of course can be a result of the approximated circular shape of the cell. (color online)}
\label{figdatafit}
\end{figure}

The analytical flow field of a circular squirmer particle can be described by a superposition of a uniform background velocity, in our case, the constant velocity of the moving cell $v_{cell}$, a Stokeslet, a stresslet and a source doublet. In \cite{Drescheretal_PRL_2010} this is used to identify typical experimental flow fields. We here consider the same approach and use the velocity field of a circular cell with center of mass $\mathbf{x}_{cm}=(0,0)^\top$ in a co-moving frame, given by
\begin{equation}
 \mathbf{v}(\mathbf{r})=-v_{cell}\mathbf{e}_1-\frac{A_{st}}{r}(\mathbf{I}+\mathbf{r}\cdot\mathbf{r})\mathbf{e}_1-\frac{A_{str}}{r^2}(1-3\left(\frac{x_1}{r}\right)^2)\mathbf{r} -\frac{A_{sd}}{r^3}\left(\frac{\mathbf{I}}{3}-\mathbf{r}\cdot\mathbf{r}\right)\mathbf{e}_1
\end{equation}
where $\mathbf{r}=\mathbf{x}/r$ is the polar axis, scaled with the distance $r=\sqrt{x_1^2+x_2^2}$, $\mathbf{e}_1$ the unity vector in $x_1$-direction and $\mathbf{I}$ the identity matrix. We prepared our numerical solution: $\mathbf{u}=(u_1-v_{cell},u_2)^T$, $\mathbf{x}=(x_1-x_{cm},x_2-y_{cm})^T$ and claim $|\mathbf{u}-\mathbf{v}| \to \min$ outside the circular cell shape with radius $R=5$ to determine $v_{cell}$, $A_{st}$, $A_{str}$ and $A_{sd}$. Table \ref{tab:opt} shows the parameters obtained from the data fit. For 1/Fa=0.5 the stresslet parameter $A_{str}$ is negative which indicates a puller like velocity profile and for 1/Fa=1 $A_{str}$ is positive, indicating a pusher like velocity profile. For 1/Fa=0.75 the data fit suggests a low puller like velocity profile. However, we should keep in mind that we compare velocity profiles of a circular and a non-circular shape. This discrepancy can be seen by analyzing the relative error $|\mathbf{u}-\mathbf{v}|/v_{cell}$ between the numerical results and the 
fitted analytical solution, see Fig. \ref{figdatafit}. The maximum of the error appears at the part of the cell, where it is compressed and does not overlap with the circular shape.
\begin{table}[H]
\centering
\begin{tabular}{|l|l|l|l|l|}
\hline
1/Fa & $v_{cell}$          & $A_{st}$               & $A_{str}$       & $A_{sd}$\\    
\hline
$0.50$ &  0.0294         &   0.0387                 &   -0.3541       &   12.5882  \\
$0.75$ &  0.0701         &   0.0872                 &   -0.1744       &   28.8854  \\
$1.00$ &  0.1089         &   0.1460                 &    0.3910       &   47.3611  \\
\hline
\end{tabular} 
\caption{Optimal parameters for background velocity $v_{cell}$, the Stokeslet $A_{st}$, the stresslet $A_{str}$ and the source doublet $A_{sd}$ obtained from a data fit with the numerical solution.}
\label{tab:opt}
\end{table}

Even if a transition from puller-like to pusher-like dynamics can be observed for increasing actin-myosin interactions, the flow characteristics are much less developed than in typical squirmer models \cite{Molinaetal_SoftMatter_2013} and are dominated by the Stokeslet contribution. Within the analytical treatment of a circular droplet in \cite{Whitfieldetal_EPJE_2014} it was found that the droplet behaves like a puller. However, for the small splay considered, the corresponding flow field is not sufficient for motility and it is the quadrupole moment that characterizes the motility mechanism, resembling the motility mechanism of a squirmer. This is consistent with our findings for low 1/Fa.

In \cite{Drescheretal_PRL_2010} the same fitting approach is used to analyze the flow topology for swimming microorganisms, such as {\it Cloamydomonas reinhardtii} and {\it Volvox carteri}. Here, the flow is also strongly dominated by the Stokeslet contribution and puller like dynamics are only mildly developed. However, for a quantitative comparison of our results with the flow fields of such microorganisms, or that of bacteria, which typically show pusher-like dynamics, more experimental data are required. It would be interesting how predictions of the considered model in 3D compare with such measured flow fields in the future.

\section{Conclusion}

We here review and extend a proposed generic model for cell motility \cite{Tjhungetal_PNAS_2012}, which is based on spontaneous symmetry breaking in active polar gels. It models the interaction of myosin and actin as the driving mechanism for motility and does not require adhesion. The model is extended to include further membrane properties, in particular bending properties, which however turn out to be of minor relevance for motility in the considered parameter regime. Detailed numerical studies are performed and convergence studies considered to demonstrate the stability of the used algorithm, which is based on a phase-field description. The results clearly indicate the independence of the physical instabilities, the splay or bend instability, which are responsible for cell motility in the considered model, and possible numerical instabilities and show the robustness of the motility mode. With this confidence in the model and the developed numerical algorithm, the results are compared with model and 
experimental data for swimming microorganisms. Within certain parameter regimes a 
transition from puller-like to pusher-like dynamics can be found for increasing myosin-actin interactions, demonstrating the generic properties of the model. A quantitative comparison with swimming microorganism is not yet possible and besides the lack of available experimental data, requires 3D simulations and probably further model extension. One possible way to extend the model is a combination of the myosin-acting interactions with the treadmilling process of acting polymerization and depolymerization, described in the introduction. However, qualitative similarities with generated flow fields of microorganisms, such as {\it Volvox carteri} could already be found. The simulated flow field as well as the measured flow field is dominated by the Stokeslet contribution. In \cite{Drescheretal_PRL_2010} it is argued that this behavior is going to have an effect on the rheology of suspensions of such microorganisms. With these properties, suspensions of our modeled cells would probably behave more like 
suspensions of sedimenting particles, as higher order moments are negligible in flow fields dominated by the Stokeslet contribution. However, if this assumption holds, or the weakly developed puller- or pusher-like dynamics in the considered model are already sufficient to observe typical phenomena in active fluids, as e.g.  phase-separation, have to be tested.

\vspace*{0.5cm}
{\bf Acknowledgement}: W.M. and A.V. acknowledge support from the German Science Foundation through grant Vo-899/11. Simulations were carried out at ZIH at TU Dresden and JSC at FZ Julich.

\appendix

\section{Thermodynamic consistency} \label{app:1}
Without the active terms the proposed system of equations is thermodynamically consistent.  To show this, we consider 
\begin{equation}
{\dot E}(\mathbf{P}, \phi, \mathbf{u})={\dot E}_\mathbf{P} + {\dot E}_S + {\dot E}_{kin} =\int \mathbf{P}^\natural \cdot \partial_{t} \mathbf{P} + \phi^\natural \partial_{t} \phi + \mathbf{u} \cdot \text{Re} \partial_{t} \mathbf{u} \dx
\end{equation}
with
\begin{align}
\partial_{t}\mathbf{P}  &= - (\mathbf{  u} \cdot  \nabla)\mathbf{P} -  \mathbf{\Omega} \cdot \mathbf{P} + \xi \mathbf{  D} \cdot \mathbf{P} - \frac{1}{\kappa}\mathbf{P}^\natural \label{eq:P_ed}\\
 \partial_t \phi &= - \nabla \cdot \left(\mathbf{u} \phi \right) + \gamma \Delta \phi^\natural \label{eq:phi_ed}\\
\text{Re}\partial_t \mathbf{u}  &= -\text{Re}(\mathbf{u}\cdot \nabla)\mathbf{u} - \nabla p + \nabla\cdot (\sigma_{viscous}+\sigma_{dist}+\sigma_{ericksen}) \label{eq:ns_ed} 
\end{align}
which yields
\begin{align*}
{\dot E}(\mathbf{P}, \phi, \mathbf{u}) &=\int \mathbf{P}^\natural\cdot (- (\mathbf{  u} \cdot  \nabla)\mathbf{P} - \mathbf{\Omega} \cdot \mathbf{P} + \xi \mathbf{  D} \cdot \mathbf{P} - \frac{1}{\kappa}\mathbf{P}^\natural) \dx \\
             & \quad + \int \phi^\natural (- \nabla \cdot \left(\mathbf{u} \phi \right) + \gamma \Delta \phi^\natural) \dx\\
	    & \quad + \int \mathbf{u} \cdot(- \text{Re}  (\mathbf{u}\cdot \nabla)\mathbf{u} - \nabla p + \nabla\cdot (\boldsymbol{\sigma}_{viscous}+\boldsymbol{\sigma}_{dist}+\boldsymbol{\sigma}_{ericksen}) )\dx \\
&=\int -\frac{1}{\kappa} |\mathbf{P}^\natural|^2 - \gamma |\nabla \phi^\natural|^2 \dx \\ 
& \qquad \quad \text{\footnotesize{(partial integration)}} \\
& \quad + \int \mathbf{u} \cdot \left( - \nabla\mathbf{P}^\top \cdot \mathbf{P}^\natural -\phi^\natural\nabla\phi  + \nabla\cdot\boldsymbol{\sigma}_{ericksen} \right) \dx \\
& \qquad \quad \text{\footnotesize{(use $\nabla \cdot \mathbf{u} = 0$)}} \\
& \quad + \int \nabla \mathbf{u} : \left( \frac{1}{2}\mathbf{P}^\natural\otimes \mathbf{P} - \frac{1}{2}\mathbf{P} \otimes\mathbf{P}^\natural + \frac{\xi}{2}\mathbf{P}^\natural\otimes \mathbf{P} + \frac{\xi}{2}\mathbf{P} \otimes\mathbf{P}^\natural - \boldsymbol{\sigma}_{dist} \right) \dx \\ 
& \qquad \quad \text{\footnotesize{(partial integration, definition for $\mathbf{\Omega}=\frac{1}{2}(\nabla \mathbf{u}^\top-\nabla \mathbf{u})$ and $\mathbf{D}=\frac{1}{2}(\nabla \mathbf{u}+\nabla \mathbf{u}^\top)$)}} \\
& \quad  + \int - |\nabla \mathbf{u}|^2 \dx \\
& \qquad \quad \text{\footnotesize{(partial integration, use $\nabla \cdot \mathbf{u} = 0$ and $\boldsymbol{\sigma}_{viscous} = \mathbf{D}$)}} \\
& \leq 0,
\end{align*}
where we have used the definition for $\nabla \cdot \boldsymbol{\sigma}_{ericksen}$ and $\boldsymbol{\sigma}_{dist}$, which show that the integrals involving these terms vanish, and the identity $\mathbf{u} \times ( \nabla \times \mathbf{u}) = \nabla (|\mathbf{u} |^2) - (\mathbf{u} \cdot \nabla) \mathbf{u}$ from which follows that $\int \mathbf{u} \cdot (- \text{Re} (\mathbf{u} \cdot \nabla) \mathbf{u}) = 0$.

\section{Numerics} \label{app:2}
The system of partial differential equations is discretized using the parallel adaptive finite element toolbox AMDiS \cite{Veyetal_CVS_2007,Witkowskietal_ACM_2015}. 

\subsection{Time discretization}
We split the time interval $I = [0,T]$ into equidistant time instants $0 = t_0 < t_1 < \ldots$ and define the time steps $\tau := t_{n+1}- t_n$. Of course, adaptive time steps may also be used. We define the discrete time derivative $d_t \cdot ^{n+1} := (\cdot ^{n+1} - \cdot ^{n})/ \tau$, where the upper index denotes the time step number and e.g. $\phi^n:=\phi(t_n)$ is the value of $\phi$ at time $t_n$. In each time step we solve:
\begin{enumerate}
\item the flow problem for $\mathbf{u}^{n+1}$ and $p^{n+1}$:
\begin{align*}
-\Delta\mathbf{u}^{n+1} +\nabla p^{n+1} &= {\phi^\natural}^n \nabla \phi^n + \nabla{\mathbf{P}^T}^n \cdot {\mathbf{P}^\natural}^n+\frac{1}{\text{Fa}} \nabla \cdot \left( \tilde\phi^n \mathbf{P}^n \otimes \mathbf{P}^n \right) +  \nonumber\\
& \quad  + \frac{1}{2} \nabla \cdot \left({\mathbf{P}^\natural}^n \otimes\mathbf{P}^n -\mathbf{P}^n\otimes {\mathbf{P}^\natural}^n \right)  \nonumber \\
& \quad + \frac{\xi}{2} \nabla \cdot \left({\mathbf{P}^\natural}^n \otimes \mathbf{P}^n + \mathbf{P}^n \otimes{\mathbf{P}^\natural}^n \right),\\
\nabla \cdot \mathbf{u}^{n+1} &=0.
\end{align*}
\item The orientation field for $\mathbf{P}^{n+1}$:
\begin{align*}
d_t\mathbf{P}^{n+1} + (\mathbf{ u}^{n+1}\cdot \nabla)\mathbf{P}^{n+1}  &= - \mathbf{\Omega}^{n+1} \cdot \mathbf{P}^{n+1} + \xi \mathbf{D}^{n+1} \cdot \mathbf{P}^{n+1} - \frac{1}{\kappa}{\mathbf{P}^\natural}^{n+1}, \\
{\mathbf{ P}^\natural}^{n+1}  &= \frac{1}{\text{Pa}}\left(-c_1\phi^n \mathbf{P}^{n+1} + c_1 (({\mathbf{P}^{n+1}})^2\mathbf{P}^{n+1})\right) \\
& \quad + \frac{1}{\text{Pa}}\left(\Delta \mathbf{P}^{n+1}+\beta \nabla  \phi^n\right),
\end{align*}
where we linearize $({\mathbf{P}^{n+1}})^2\mathbf{P}^{n+1}=({\mathbf{P}^n})^2 \mathbf{P}^{n+1} +2 (\mathbf{P}^{n}\otimes\mathbf{P}^{n})\mathbf{P}^{n+1} -2({\mathbf{P}^n})^2 \mathbf{P}^{n}$.
\item The phase field evolution for $\phi^{n+1}, \mu^{n+1}, \psi^{n+1}$:
\begin{align*}
d_t\phi^{n+1} + \nabla \cdot \left(\mathbf{ u}^{n+1}\phi^{n+1}\right) &= \gamma\Delta {\phi^\natural}^{n+1},\label{eq:phase_num} \\
{\phi^\natural}^{n+1} &= \frac{1}{\text{Be}}\psi^{n+1}  -\frac{1}{\text{Ca}} \mu^{n+1} \\
& \quad -\frac{1}{\text{Pa}}(c_1 |\mathbf{P}^{n+1}|^2+\beta \nabla\cdot\mathbf{P}^{n+1}), \\
\mu^{n+1}  &= \eps\Delta\phi^{n+1}-\frac{1}{\eps}(({\phi^{n+1}})^2-1)\phi^{n+1},\\
\psi^{n+1} &= \Delta \mu^{n+1}-\frac{1}{\eps^2}(3({\phi^{n+1}})^2-1) \mu^{n+1},
\end{align*}
where we again linearize the non-linear terms by a Taylor expansion of order one, e.g. $((\phi^{n+1})^2 - 1) \phi^{n+1}=((\phi^n)^2-1)\phi^n+(3{(\phi^n)}^2-1)(\phi^{n+1}-\phi^n)$.
\end{enumerate}

\subsection{Fully discrete finite element scheme}
The fully discrete finite element scheme follows in a straight forward manner. A $P^2/P^1$ Taylor-Hood element is used for the Stokes problem, all other quantities are discretized in space using $P^2$ elements. The obtained linear system, for which the direct unsymmetric multifrontal method UMFPACK is used, is solved in each time step. We use an adaptively refined triangular mesh $\mathcal{T}_h$ with a high resolution along the cell membrane to guarantee at least five grid points across the diffuse interface as well as a high resolution within the cytoplasm to appropriately resolve the orientation field. The criteria to refine or coarsen the mesh is purely geometric and related to the phase field variable $\phi$. Due to the use of adaptivity, we need to interpolate the old solution defined on ${\cal{T}}_h^n$ onto the new mesh ${\cal{T}}_h^{n+1}$. To do this without violating the conservation of cell volume, we solve $\langle \phi^{n,old}, \theta\rangle = \langle \phi^{n,new}, \theta \rangle$ in every 
adaption step, with $\theta$ and $\phi^{n,new}$ defined on ${\cal{T}}_h^{n+1}$ and $\phi^{n,old}$ on ${\cal{T}}_h^{n}$. We use a multi-mesh strategy \cite{Voigtetal_JCS_2012} to virtually integrate the first term on the finest common mesh ${\cal{T}}_h^{n} \cup {\cal{T}}_h^{n+1}$, which guarantees a constant cell volume as long as time steps are appropriately chosen. We require the interface not to propagate over a whole element within one time step. With this restriction, all numerical tests show that $\int_\Omega \phi \; d \mathbf{x}$ is conserved. 

\bibliography{literature}
\bibliographystyle{vancouver}

\end{document}